# A texture-based framework for foundational ultrasound models

Tal Grutman, Carmel Shinar, and Tali Ilovitsh, Senior *Member, IEEE*

*Abstract*—Ultrasound is the most widely used medical imaging modality, yet the images it produces are fundamentally unique, arising from tissue-dependent scattering, reflection, and speed-of-sound variations that produce a constrained set of characteristic textures that differ markedly from natural-image statistics. These acoustically driven patterns make ultrasound challenging for algorithms originally designed for natural images. To bridge this gap, the field has increasingly turned to foundation models, hoping to leverage their generalization capabilities. However, these models often falter in ultrasound applications because they are not designed for ultrasound physics, they are merely trained on ultrasound data. Therefore, it is essential to integrate ultrasound-specific domain knowledge into established learning frameworks. We achieve this by reformulating self-supervised learning as a texture-analysis problem, introducing texture ultrasound semantic analysis (TUSA). Using TUSA, models learn to leverage highly scalable contrastive methods to extract true domain-specific representations directly from simple B-mode images. We train a TUSA model on a combination of open-source, simulated, and in vivo data. The latent space is compared to several larger foundation models, demonstrating that our approach gives TUSA models better generalizability for difficult downstream tasks on unique online datasets as well as a clinical eye dataset collected for this study. Our model achieves higher accuracy in detecting COVID (70%), spinal hematoma (100%) and vitreous hemorrhage (97%) and correlates more closely with quantitative parameters like liver steatosis ($r = 0.83$), ejection fraction ($r = 0.63$), and oxygen saturation ($r = 0.38$). We open-source the model weights and training script: https://github.com/talg2324/tusa.

*Index Terms*—Contrastive learning, ultrasound foundation models, deep learning, ultrasound.

This work was supported in part by the Israel Science Foundation under Grant 192/22, in part by an ERC StG under Grant 101041118 (NanoBubbleBrain), in part by the Israel Cancer Research Fund (grant number 1286686), and in part by the Nicholas and Elizabeth Slezak Super Center for Cardiac Research and Biomedical Engineering at Tel Aviv University. Tal Grutman and Tali Ilovitsh are with the School of Biomedical Engineering, Tel-Aviv University, Tel Aviv, 6997801, Israel (e-mail: ilovitsh@tauex.tau.ac.il). Carmel Shinar is with the Tel-Aviv Sourasky Medical Center and School of Medicine, Tel-Aviv University, Tel Aviv, 6997801, Israel.

## I. INTRODUCTION

D<small>EEP</small> learning has revolutionized computer vision, with practical applications in both the real world and medical imaging. Ultrasound has benefited greatly from the capability of deep learning models to adapt themselves to data, while classical vision algorithms often struggle with its lower resolution, contrast, and signal-to-noise ratio. Many recent works in ultrasound utilize deep learning for various applications including holography[1], beamforming[2], volume representation [3], and more. Despite this, the scale of data that tends to be used in frontier deep learning algorithms [4] is harder to acquire in ultrasound, where supervised labels often undergo heavy scrutiny by multiple specialists [5], [6] due to the vastly higher consequences for error in pathology detection common to clinical imaging.

Ultrasound is fundamentally different in the way images are created compared to other medical imaging modalities, posing challenges for natural-image-based algorithms. In a typical ultrasound setup, a transducer array transmits a sound wave into the target medium. As the wave propagates, it is scattered and reflected by the target anatomy due to its geometry and composition. The transducer elements record the returned echoes, and the ultrasound image is reconstructed from the time-of-flight of each echo returning to the transducer and from the echo amplitudes, which map into a gray-scale image. The trajectory of the sound waves in the medium is determined mainly by the local speed of sound and by the scattering and reflection coefficients of the tissue. These are physiological parameters that vary locally. The gray-scale levels in the image form texture patterns that arise from the spatial distribution of these acoustic properties. Although acoustic parameters vary among people, organs, and pathologies, they fall within a constrained physiological range in the human body. Consequently, despite the biological diversity of human tissues, ultrasound images contain a limited set of characteristic textures. Organ boundaries and bone typically appear white, soft tissues display grainy speckle patterns, and fluid appears black. This explains why additional context is needed for interpretation: a dark anechoic region in the heart represents blood, whereas the same appearance in obstetric imaging indicates amniotic fluid. Furthermore, the detected sound wave signals are encoded into image pixels on a logarithmic scale to accommodate the large dynamic range between weak reflections and strong scattering. These combined phenomena produce images that differ substantially from classical natural images and from other radiological modalities.

Despite these differences, the visual content in many medical imaging modalities is constrained to a limited and well-defined set of human tissues. This constraint has motivated the development of modality-agnostic models like RadImageNet[7] and MedSAM[8], which leverage shared anatomical structures across modalities to learn transferable representation. More recently, the field has shifted toward "foundation models," large models trained on massive datasets using unsupervised or self-supervised approaches[9]. With growing dataset and model size, foundation models learn to generalize in a way that exceeds the capacity of their training data. Such models, although costly to train, are very useful in downstream applications. Foundation models can be used for many applications including out-of-the-box (zero-shot) usage on specialized datasets, extraction of useful features for downstream analysis, or fine-tuning on a more specific task.

These use-cases are immensely practical in medical imaging and especially ultrasound, where high quality labelled datasets are scarce. Several foundation models have been developed and shared for both general medical diagnostics [8], [10] and ultrasound specifically [11], [12], [13]. They are considered a "democratization" of automated ultrasound analysis, as they provide lay users with a model that has instilled in it the knowledge of potentially ultra-expensive datasets and GPU hours. However, these models leave much to be desired in generalizability across ultrasound domains because they are not designed for ultrasound images, just trained on ultrasound data. For example, UltraSam[12] struggled to adapt to datasets acquired with unique probes that were not represented in the training data. This highlights the need for an ultrasound-first design that is based on the underlying imaging principles rather than being implicitly extracted from the dataset. Nevertheless, foundation models continue to be trained in generalist frameworks because specialist knowledge does not scale well.

Thus, it is crucial to integrate ultrasound domain knowledge into well-studied learning frameworks like masked autoencoding [14] and SimCLR [15] that can take advantage of data augmentations to create something from nothing: an unlabeled image can be compared to an augmented version of itself. Such techniques have shown considerable gains in performance on natural images and remain a widely used pattern for model pre-training even outside the scope of foundation models[16], [17]. These techniques have begun to be adapted in ultrasound projects like Ultrasound Foundation Model (USFM), which used self-supervised learning to train from B-mode images in a masked autoencoding framework [13]. In this work, we will use gray-scale texture patterns as an intermediary between classical self-supervised learning and ultrasound-inspired deep learning architectures. By incorporating ultrasound-specific texture analysis into an autoencoding framework, we hypothesize that ultrasound-specific intuition can be embedded into a deep learning model, harnessing the texture information that is common to various areas of the body, an approach that is being adopted by other modalities as well [18]. By formulating self-supervised learning as a texture analysis problem, we can take advantage of the highly scalable contrastive frameworks that have shown mainstream success to create real domain-specific knowledge from simple B-mode images. Our framework is based on the intuition that human professionals assess ultrasound images by first identifying macro-level structures in gray-scale images- often unified by their textures- before solving the task at hand.

We first proposed this training framework, termed texture ultrasound semantic analysis (TUSA) in [19], and have considerably refined and improved it with more data and robust validation studies in this paper, which will describe the integration of texture analysis into self-supervised learning before analyzing the latent space of a TUSA model compared to several well-known foundation models like MedSAM[8], UltraSam[12], and USFM[13], highlighting the importance of scalable ultrasound-based information in the extraction of features from B-mode images

## II. METHODS

Our TUSA model learns texture by splitting the task of image autoencoding into two parts. The first task is a segmentation problem. Given $K$ texture channels, an encoder-decoder network of choice is used to segment each B-mode pixel to each of the $K$ texture channels. In the second phase, each channel is convolved with its own learnable convolutional kernel, before the channels are added together to reconstruct the original image intensity. This separation of tasks essentially redefines autoencoding as a separation of shapes from textures in a way that can be convolved together to reproduce the original image (Fig. 1).

During training, TUSA learns to complete both tasks in sequence. The limited number of allowed texture kernels forces TUSA to identify textures that are consistent across various organs, taking advantage of recurring intensity patterns in B-mode to learn to segment the different organs. This domain-specific solution is still an autoencoding problem at heart that can be trained without labels, making it highly scalable.

Once trained, TUSA can be employed for any of the use-cases described previously. The encoder can be detached from the decoder to encode texture-based latent representations of B-mode images, providing valuable insights into the image content. The segmented texture image can be filtered to isolate specific textures or fine-tuned to a particular segmentation task.

In this paper, we will focus on latent-space knowledge encoded by foundation models. Three methods will be tested for probing the feature extraction capability of foundation models: visualization, soft binary classification, and downstream regression.

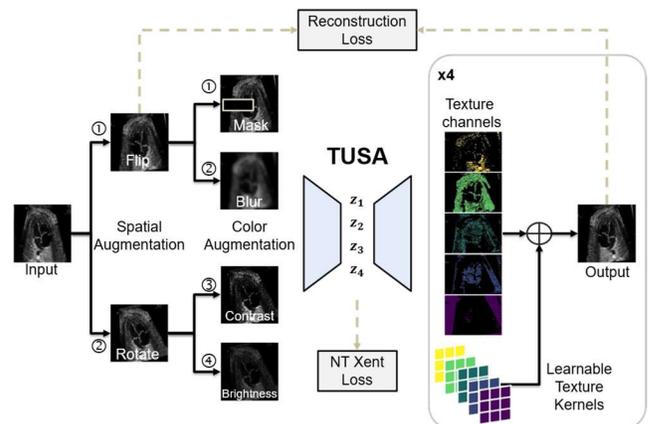

Fig. 1. Deep autoencoding reformulated as a two step problem. First images are segmented to produce $K$ texture channels. In the second stage, each channel is convolved with a unique, learnable kernel to reproduce input intensity.

### A. Implementation and Training
#### 1) Model Implementation
We used sliding window U-Net Transformer (SwinUNETR)

[20], a recent segmentation architecture that integrates convolutional and attention-based information to extract global and local features. The logits of this model undergo Sparsemax [21] activation before the reconstruction step, encouraging TUSA to select a specific texture channel for each pixel in the image. In the reconstruction phase, a depth-wise separable convolution convolves each channel with a unique, learnable kernel, before the channel dimension is squashed by a 1x1 kernel with tanh activation to reconstruct the B-mode intensity. We used $K=5$ texture kernels after empirically noting that additional kernels produced redundant information at the end of the segmentation phase.

TUSA is trained according to the SimCLR framework with several well-established loss functions focusing on reconstruction in addition to self-supervised penalties and regularization. An input image is spatially augmented randomly twice. Each of the spatial augmentations is then augmented with random color augmentations (erasing, brightness, contrast, Gaussian blur). During a model forward pass on these four augmentations, we calculate the normalized temperature cross-entropy (NT-Xent) loss [15] between the encodings of the four images and the rest of the images in the batch. We expect that our model will develop robustness to simple addition and multiplicative intensity changes since it is supposed to identify macro-level texture patterns, so mean squared error is computed between logits resulting from color augmentations of the same spatial augmentation. Once we receive the four output reconstructions for this image, the reconstruction loss, which is composed of L1 loss, structural similarity (SSIM) [22] and learned perceptual image patch similarity (LPIPS) [23] with RadImageNet backbone [7] is calculated. This training pipeline is described in Fig. 1.

Lastly, we encourage the model to maximize its usage of the various texture channels by weakly penalizing the loss function with the negative channel-wise entropy of its output. This encourages TUSA to maximize its usage of the different textures within the batch and helps to protect it from learning redundant features.

TUSA was trained on a remote Kubernetes cluster with the Run:ai (Tel-Aviv, Israel) resource manager on an NVIDIA RTX A5000 GPU (NVIDIA Corporation, Santa Clara, CA, USA). We used the PyTorch [24] framework and the MONAI [25] project for model implementation and LPIPS loss. The model was trained for 1k epochs of 10k images using mixed precision. We used the AdamW [26] optimizer and a batch size of 32 with an initial learning rate of $10^{-4}$ and cosine annealing down to $10^{-6}$ over the course of training. Major regions of the body were sampled uniformly so that over-represented organs (most notably the echocardiogram datasets) would not dominate the learned textures (Fig. 2). A similar sampling technique was used to train USFM [13], and it is crucial in order to differentiate between the prominence of a particular texture from its availability in open-source datasets. TUSA candidates were manually reviewed on a validation set of images consisting of a single frame from each of the various body regions included in the training data.

### 2) Training and Evaluation Data

Our model is trained and evaluated on several types of datasets. First, we utilize publicly available ultrasound datasets from diverse organs and hardware for both training and evaluation. The training data is supplemented with k-wave [27], [28] phantoms, in-silico phantoms, and in-vivo breast cancer tumors in mice collected in a prior work [3]. To supplement our evaluation data, we add an additional clinical dataset of eye images collected in the clinic. In total, the training data is composed of 100k B-mode images from 9 areas in the body, as well as a small subset of supplementary data collected in a previous work in our lab (breast cancer tumors in mice and simulated phantoms). The various training datasets are listed in Table I.

Table I. Training dataset and organ group assignment

| Organ Group | Dataset | Number of Images |
|---|---|---|
| Abdominal | AbdominalUS [29] | 615 |
| Abdominal | LiverTumors [30] | 635 |
| Abdominal | OpenKidney [6] | 535 |
| Breast | BrEaST-Lesions_USG [31] | 270 |
| Breast | BUS-UCLM [32] | 683 |
| Breast | BUSBRA [33] | 1,873 |
| Breast | BUSI [34] | 780 |
| Cardiac | CAMUS [35] | 21,029 |
| Cardiac | TED [36] | 4,524 |
| Cardiac | UnityImaging [37] | 7,494 |
| Fetal | Fetal Abdominal Structures Segmentation Dataset Using Ultrasonic Images [38] | 1,587 |
| Fetal | Fetal Planes [39] | 12,336 |
| Fetal | PSFHS [40] | 4,064 |
| Follicle | OTU_2d [41] | 1,469 |
| Knee | JoCoHS [42] | 9,575 |
| Lung | Covid19 Ultrasound [43] | 19,955 |
| Musculoskeletal | deepMTJ [44] | 2,683 |
| Musculoskeletal | FALLMUD [45] | 504 |
| Neck | Common Carotid Artery Ultrasound Images [46] | 1,100 |
| Neck | DDTI [47] | 480 |
| Neck | Thyroid Dataset [48] | 7,056 |
| Lab | Mouse breast cancer [3] | 720 |
| Lab | k-wave phantom [27], [28] | 432 |

All images are converted to grayscale, resized to a uniform resolution of $128^2$, and cast to the range of [-1, 1]. We evaluate our model on more recent and difficult datasets from parts of the body that are typically underrepresented in open-source data including our proprietary dataset of eye ultrasounds. We use these challenging datasets to investigate the generalizability of models outside of their training distribution. The evaluation datasets are described in Table II.

Table II. Evaluation datasets and tasks.

| Organ Group | Dataset | Tasks |
|---|---|---|
| Lung | COVID-BLUeS [49] | Visualization, binary classification, downstream regression |
| Spinal cord | Ultrasound spinal cord injury [5] | Visualization, binary classification |
| Eye | Vitreous hemorrhage | Visualization, binary classification |
| Cardiac | EchoNet-Dynamic[50] | Downstream regression |
| Abdominal | Liver Steatosis[51] | Downstream regression |

### 3) In-human Eye Pathology Dataset

US B-mode images were acquired using a commercially available, FDA-cleared ocular ultrasound system equipped with 15 MHz and 20 MHz probes. All imaging was performed under routine clinical conditions in an active ophthalmologic practice using an ABSOLU ultrasound system (Quantel Medical, Cournon-d'Auvergne, France). The study included eyes from 54 patients with recurring visits, with each eye associated with one clinical case at each visit. A total of 446 images were acquired, corresponding to 118 clinical cases, of which 29 cases contained vitreous hemorrhage. Hemorrhages varied in etiology and were associated with several underlying ocular pathologies and a wide range of severities, from dense opacities obscuring anatomical landmarks to minimal or subclinical presentations. Clinical assessment was conducted via standard ophthalmologic examination, and findings were compared with those obtained through ultrasonography. In certain cases, hemorrhages were identified only through clinical examination and not visualized on ultrasound, and vice versa. Eyes without pathological vitreous opacities served as controls.

All imaging procedures followed the protocol approved by the Tel Aviv Sourasky Medical Center and Tel Aviv University Institutional Review Board (TLV-0769-25).

### 4) Downstream Tasks

We compared the latent space of TUSA to that of established and specialist foundation models like MedSAM, UltraSam, and USFM. It is important to note that each of these models contains an 86M parameter vit-b encoder, compared to TUSA's 9M parameter SwinViT. Identical images were used for all tests, but images were resized and rescaled according to the specifications of each model.

Three downstream tasks were evaluated (Table II). To focus on the information extracted by each model, we focused on tasks where the model encoder was frozen and not fine-tuned. First, latent space visualization was performed using uniform manifold approximation and projection (UMAP) [52]. A random sample of 200 images was selected from each organ group. Each model encoded the visualization set to latent vectors on which the unsupervised UMAP algorithm was employed to create a 2D visualization. Next, soft latent space classifiers were fit using the linear support vector classifier (SVC) to classify COVID-19 in lung ultrasound, spinal hematoma in spinal cord images, and vitreous hemorrhage in eye ultrasound. The UMAP projection of each model is also shown for these datasets to assess whether enough information is present to easily separate the pathology class. Finally, we tested the models on challenging regression tasks. Frozen model encoders were used to encode a series of images from a single patient. A small downstream network consisting of several linear layers followed by a temporal attention mechanism was trained to predict ejection fraction from cardiac embeddings, liver steatosis from abdominal embeddings, and oxygen saturation from lung ultrasound embeddings.

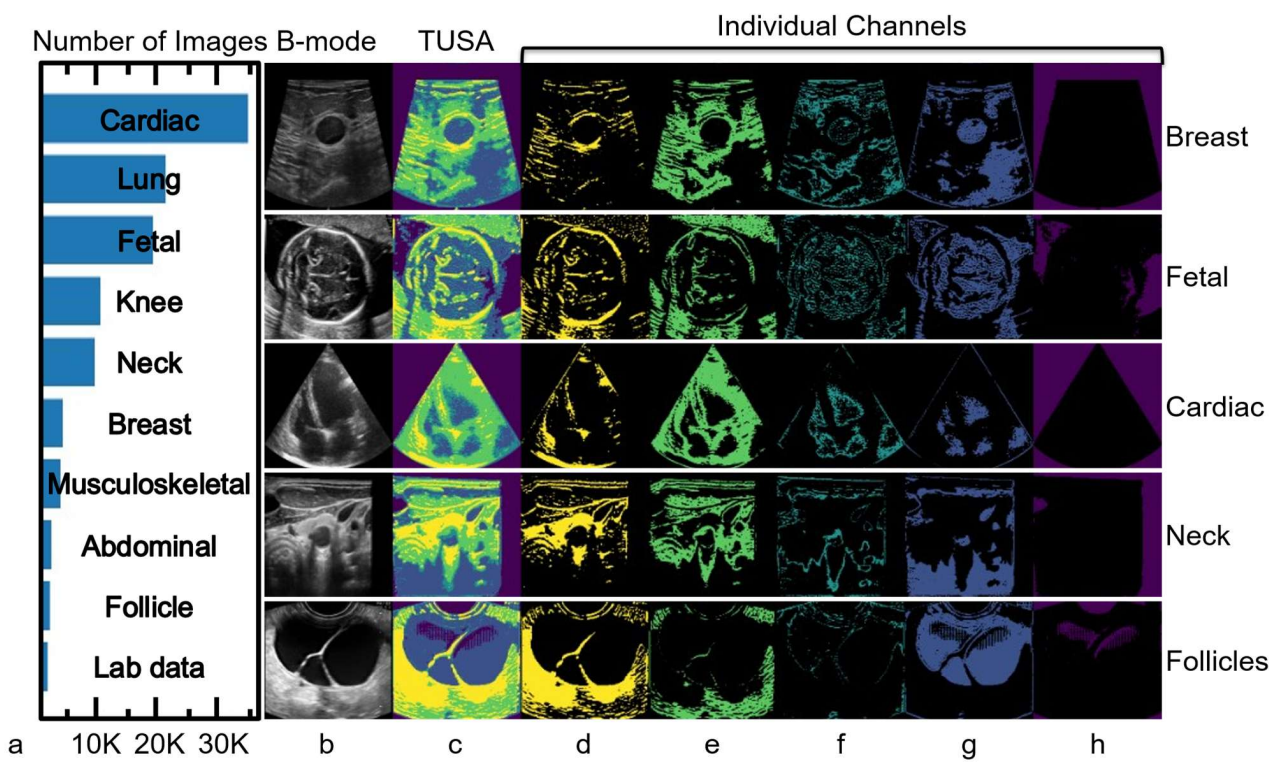

Fig. 2. Organ distribution of the training data ordered by prominence (a). Sample images are presented in standard B-mode (b) alongside the TUSA texture map (c). Individual channels (d-h) a are presented in decreasing order of echogenicity from left (most echoic) to right (anechoic).

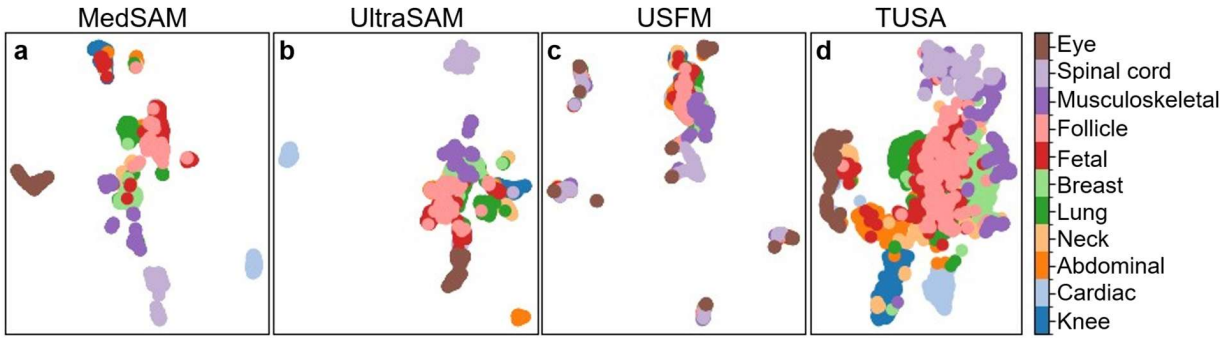

Fig. 3. Latent space visualization with UMAP. 200 images sampled from each organ group including the unique eye and spinal cord data were encoded by each of the foundation models as well as TUSA.

Code for dataset creation, training, and evaluation along with a pre-trained TUSA model is available at https://github.com/talg2324/tusa.

## III. Results

### 1) TUSA Model

Our trained TUSA model was evaluated on ultrasound images from a variety of organs (Fig. 2b-h). Here, representative images from five different organs (breast, fetal, cardiac, neck, and follicles) are presented in standard grayscale B-mode (Fig. 2b) alongside their TUSA decomposition (Fig. 2c), which is the superposition of the individual texture channels (Fig. 2d-h). When examining these channels individually, a consistent pattern emerges. The yellow channel highlights bones, muscle fibers, and strong tissue boundaries, which appear as the most echogenic structures (Fig. 2d). The green channel corresponds to the majority of soft tissues (Fig. 2e), while the cyan channel captures speckle patterns within liquids and weak tissue boundaries (Fig. 2f). The blue channel represents fluids, and the purple channel reflects the background, occasionally appearing in very homogeneous liquids such as amniotic fluid (Fig. 2f).

### 2) Visualization of the Latent Space

The UMAP algorithm is used to gain qualitative intuition into the generalizability of a model based on its ability to map similar data points together in latent space. The UMAP projection of 200 randomly sampled images from each organ group is plotted (Fig. 3). MedSAM and UltraSam (Fig. 3a,b) are especially good at separating cardiac data in latent space but struggle to differentiate other organs. USFM does not provide useful latent separation on our data (Fig. 3c). Our TUSA model (Fig. 3d) extracts a highly structured latent space, which maintains separation of different regions in the body while clustering together images from the same region.

Table III. Soft binary classifier results on evaluation datasets.

| Dataset | Model | Accuracy | F1 | Specificity | Sensitivity | AUC |
|---|---|---|---|---|---|---|
| COVID | MedSAM | **69.8** | **70.8** | **70** | **69.7** | 71 |
| | UltraSam | 60.3 | 61.5 | 60 | 60.6 | 56.2 |
| | USFM | 62 | 64.7 | 56.7 | 66.7 | 65.4 |
| | TUSA | **69.8** | **70.8** | **70** | **69.7** | **75.1** |
| Spine | MedSAM | **99.8** | **99.9** | **99.4** | **100** | **100** |
| | UltraSam | 85.7 | 89.5 | 99.4 | 81.2 | 99.1 |
| | USFM | 98.3 | 98.9 | 95.8 | 99.2 | 99.7 |
| | TUSA | **99.8** | **99.9** | **99.4** | **100** | **100** |
| Eye | MedSAM | 89.8 | 81.8 | 88.8 | 93.1 | 93.4 |
| | UltraSam | 93.2 | 87.5 | 92.1 | 96.6 | 97.8 |
| | USFM | 90.7 | 82.5 | 91 | 89.7 | 92.4 |
| | TUSA | **95.8** | **92.1** | **94.4** | **100** | **99.5** |

Focusing on the evaluation data, binary classification problems are individually analyzed following UMAP projection (Fig. 4). The projection of USFM is not particularly informative for the tasks in our dataset, while TUSA and MedSAM provide the best separability of classes. This is in line with our hypothesis that generalizability is usually achieved by bigger models with more data or specialist models with scalable domain knowledge. All models struggled with the COVID-BLUeS dataset, which is consistent with the report of the authors of the data that COVID detection from a lung ultrasound is a very challenging task. In the case of our eye dataset, all of the models provide some degree of separation between healthy eyes and vitreous hemorrhage, but our model is the only one that creates tight and separable clusters, implying high separability of classes in latent space. This is in line with the texture-based diagnosis of vitreous hemorrhage.

### 3) Soft Binary Classification

To provide a more quantitative measure of the latent space information encoded by each model, we fit a class-balance weighted Linear SVC model to each of the datasets in Fig. 4. For each model, the SVC parameters were independently optimized using a grid search to optimize for accuracy. In the COVID dataset, we randomly sampled 20 frames from each patient and calculated the average latent vector per-patient, then performed 5 cross-validation splits at the patient-level, each time evaluating the SVC model on the excluded split. This was not necessary for the spinal cord hematoma data which was already split into a training and testing set by the authors. In our eye dataset, we again calculated the average latent vector per-case with 15 cross-validation splits at the patient level.

The accuracy, F1 score, specificity, sensitivity, and area under the curve (AUC) are reported for each dataset (Table III). In the COVID and spine datasets, the TUSA and MedSAM models achieve very similar results, generalizing well to the new data. However, the best result in eye hemorrhage detection is achieved by TUSA.

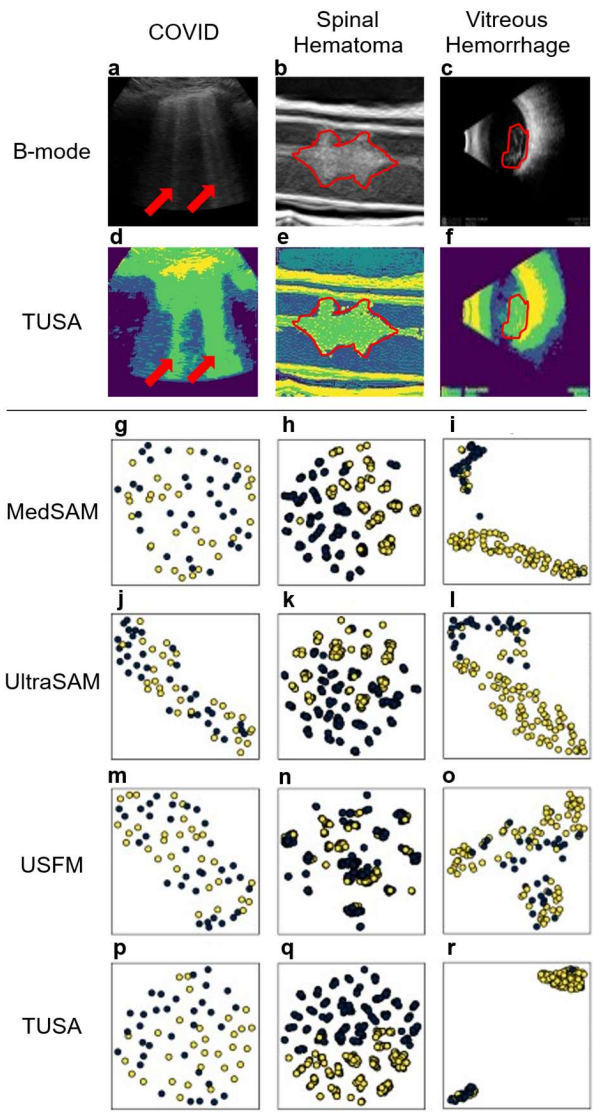

Fig. 4. Visualizing binary problems with UMAP. An input B-mode image is shown for each of the pathologies (a-c) with the pathology marked in red. Corresponding TUSA texture maps are shown for reference (d-f). UMAP projections of the binary classification datasets are displayed for each model: MedSAM (g-i), UltraSam (j-l), USFM (m-o) and TUSA(p-r). An informative latent space will show good separation between the normal class (yellow) and the pathology class (blue).

*4) Downstream Regression*

Our final evaluation was to train a downstream latent space model for predicting quantitative ultrasound metrics. This is especially difficult because it requires a latent space rich in relevant information to quantify physiological parameters, not just classify them.

We analyze the performance of such downstream modeling on fatty liver disease (Fig. 5). In healthy liver, B-mode images show strong echo from the echogenic complex of the kidney relative to the liver (Fig. 5a). Corresponding TUSA channels show strong speckle in the kidney (yellow channel) and regular soft tissue (green channel) texture in the liver (Fig. 5b). In fatty disease, the liver becomes highly echogenic, and the kidney's echogenicity is hidden because the sound wave is mostly reflected (Fig. 5c). This is detectable in the TUSA image where the TUSA image now detects high speckling in the liver (Fig. 5d), which is now segmented as the yellow channel. Such fine changes in intensity and speckling patterns are undetected by segmentation-based foundation models like MedSAM (Fig. 5e) and UltraSam (Fig. 5f), and the latent vectors encoded by those models struggle to provide meaningful quantification of liver steatosis. USFM (Fig. 5g) and TUSA (Fig. 5h), which are trained with contrastive learning schemes, are able to encode meaningful information for quantifying liver fattiness.

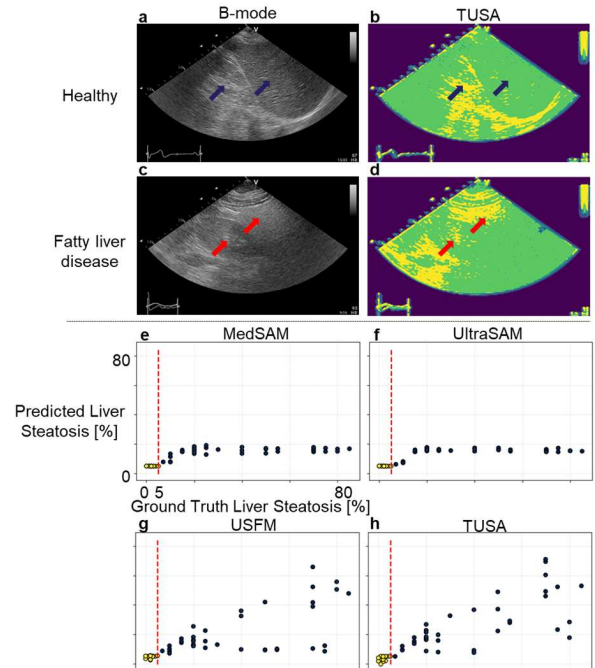

Fig. 5. Downstream regression performance in fatty liver disease. Comparison of fatty (a-b) and non-fatty (c-d) liver patients in B-mode and TUSA channels. Healthy speckling is marked in blue arrows, with fatty disease in red. Regression plots show the true label on the x axis and the model prediction on the y axis for MedSAM (e), UltraSam (f), USFM (g) and TUSA (h). Axes are identical among plots.

In the liver steatosis dataset we used all 10 available frames per patient with leave-one-out training as described by the authors, while for the EchoNet-Dynamic dataset, we selected five linearly spaced frames between systole and diastole according to the authors' annotations and used the authors' split of training and testing data. In COVID-BLUeS, we randomly sampled 20 frames from each patient and trained with 10 cross-validation splits, each time evaluating the model on 10% of the data. For each dataset, we report the pearson correlation $r$, mean absolute error (MAE), and AUC for a common regression-based classification (Table IV). In the detection of liver steatosis, the authors of the data defined 5% as the threshold for non-alcoholic fatty liver, while in EchoNet-Dynamic, the authors defined an ejection fraction threshold of 40% as separating between low-EF and normal patients. In the COVID-BLUeS dataset, we trained models to predict the blood oxygen saturation level with a threshold of 95% saturation for healthy patients, as low blood oxygen can be associated with B-mode artifacts in COVID patients [53]. We excluded the UltraSam model from the EchoNet-Dynamic evaluation as left ventricle segmentation from EchoNet-Dynamic is part of the UltraSam training data. Each latent-space model was trained ten times on each dataset with different random initializations. UltraSam and MedSAM struggled to extract any meaningful information from the latent space most likely because they are segmentation oriented models that struggle to provide non-segmentation related information without additional training. USFM does achieve meaningful downstream results on these difficult tasks but is noticeably outperformed by TUSA.

Table IV. Downstream regression results on evaluation datasets.

| Dataset | Model | r | MAE | AUC |
|---|---|---|---|---|
| Liver Steatosis | MedSAM | 0.72 ± 0.01 | 17.65 ± 0.08 | **1.00 ± 0.00** |
| | UltraSam | 0.72 ± 0.01 | 17.64 ± 0.08 | **1.00 ± 0.00** |
| | USFM | 0.67 ± 0.04 | 13.50 ± 0.56 | **1.00 ± 0.00** |
| | TUSA | **0.83 ± 0.01** | **10.03 ± 0.19** | 0.99 ± 0.00 |
| EchoNet-Dynamic | MedSAM | 0.19 ± 0.03 | 8.62 ± 0.09 | 0.64 ± 0.02 |
| | USFM | 0.57 ± 0.01 | 7.34 ± 0.06 | 0.83 ± 0.01 |
| | TUSA | **0.63 ± 0.00** | **6.95 ± 0.08** | **0.87 ± 0.01** |
| COVID-BLUeS | MedSAM | 0.37 ± 0.02 | 3.76 ± 0.01 | 0.70 ± 0.02 |
| | UltraSam | 0.29 ± 0.01 | 3.70 ± 0.02 | 0.70 ± 0.01 |
| | USFM | 0.23 ± 0.03 | 3.83 ± 0.04 | 0.64 ± 0.02 |
| | TUSA | **0.38 ± 0.03** | **3.65 ± 0.04** | **0.71 ± 0.01** |

## IV. DISCUSSION

In this work, we introduced a new self-supervised training framework for deep ultrasound models that takes advantage of scalable domain knowledge. Our TUSA framework utilizes data augmentation based on SimCLR to learn texture kernels corresponding to common ultrasound image textures while learning to segment how the kernels describe a given input image. The model essentially separates textures from anatomical structures, analyzing images as we expect humans to. To assess whether this framework is beneficial in downstream ultrasound applications, we evaluated the generalizability of a model trained in this methodology and well-established foundational models in ultrasound, seeking to highlight the tradeoff between more data and domain specialization. This is highly notable in the two edge cases of MedSAM, the model trained on the most training data (including but not specific to ultrasound), and our TUSA model, a considerably smaller specialist model trained on just 100K images. These two models outperformed the others in all of our latent space tests, providing better separation of organ groups after UMAP projection and better results in binary classification. In many examples, our TUSA model outperformed MedSAM, suggesting that a combined approach of scalable specialization with generalist frameworks like SimCLR can outperform training routines that were invented for natural images relying on vast data resources. Thus, although specialist models like UltraSam and USFM could not compete with the generalizability that MedSAM acquired from its enormous dataset, our model was able to do so by scaling specialist knowledge, which does not necessarily need to replace the data-first approach but provides an auxiliary pipeline for generalizability in foundation models.

From the latent space, it is clear that cardiac data was extensively used in training the SAM models, as they differentiate cardiac data particularly well from other ultrasound data, and this is in line with the vast availability of echocardiogram data relative to other organ groups. Similarly, the three foundation models do well with breast tumors, especially USFM whose dataset is composed primarily of breast ultrasounds. The SAM models generalize well to the unseen spinal cord and eye datasets as well as the lesser-used knee ultrasound dataset, which USFM struggles to differentiate, suggesting that USFM may not be capable of generalization to unseen organ groups. It is also interesting to see that in addition to excellent separation of the organ groups our TUSA model splits the musculoskeletal dataset into two parts, highlighting some contrast between the two datasets that compose this organ group. The latent space of the evaluation datasets shows that texture-based analysis is a better indicator of these pathologies (COVID19, spinal hematoma, and vitreous hemorrhage in the eye) than natural image-based analysis (Fig. 4). This is intuitive because the datasets on which these foundation models were trained does not include data remotely similar to the unique spinal cord and eye datasets, highlighting the importance of scalable generalizability in foundational ultrasound models. We further validated this theory in a more quantitative way using SVC models (Table III), showing that the TUSA encoding provides the most separable latent space for binary classification. Here we saw again that the best results are achieved by using the most data (SAM models) or modality specialization (TUSA). Finally, we showed that the texture knowledge ingrained in TUSA's latent space is the most applicable to real-world downstream tasks without significant additional training (a downstream MLP with pre-encoded latent space takes less than 10 minutes to train on even modest GPUs), especially in difficult and unique tasks that are different from classical ultrasound datasets available online (Table IV). This was most notable in the context of liver steatosis and ejection fraction prediction, where TUSA outscored the other models considerably.

There are several important implications to the success of our TUSA model. First, our scalable approach can be utilized to continue to push the boundaries of big specialist foundation models like UltraSam and USFM. However, our model also suggests that a smaller model can provide much of the same benefit and often outperform a foundation model when generalizing within the domain, depending on how much of an advantage there is to the usage of texture information.

Our model is an ultrasound-first model and unlike the much larger MedSAM is likely not optimal for other applications in radiology. However, the concept of learnable textures could be relevant for modality-specific foundation models by taking advantage of modality-specific features[54]. There are a few other use-cases we can suggest with TUSA. We studied the extraction of texture information in latent space and our model can be used as a small image encoder with good generalizability. In addition, the output channels received from TUSA's segmentation step can be seen as a type of component analysis similar to PCA, allowing for easy interaction with or filtering of textures not relevant to a particular analysis. Future work can analyze TUSA images like those in Fig. 2, 4, and 5 as a prior when attempting to estimate acoustic parameters. For example, prior knowledge of the speed-of-sound has been shown to greatly improve absolute estimation[55]. Alternatively, individual TUSA channels can separate relevant structures in different types of images (Fig. 2), and downstream applications could use TUSA as a starting point for morphological segmentation algorithms.

Despite showing promise in downstream applications, there is much room to improve our model. Although we empirically selected $K=5$ texture kernels, it is possible that better performance can be achieved using multi-scale kernels at higher resolution. As the number of textures used by the model rises, it would also be beneficial to create a mechanism for combining similar texture kernels and lowering redundancy. Our choice to use a smaller model with relatively modest image resolution was due to hardware constraints, and it is possible that a larger dataset and GPU could be used to train an even more generalist model. In addition, our model is trained to rely on reference textures that are common to the anatomies of the training data and would presumably struggle to extend to textures not included- for example during imaging with ultrasound contrast agents.

Nonetheless, the learned texture kernels extracted from our model provide important information on textures common to the body, and our pipeline creates an inherent separation of anatomical shapes from B-mode texture, potentially enabling the creation of synthetic images from shapes of a target anatomy. Most importantly, however, our model provides excellent ultrasound generalizability with only 22M parameters and little specialized code, relying mostly on the standard implementation of the MONAI library and allowing for easy plug-and-play usage and fine-tuning for downstream and zero-shot scenarios. The same cannot be said for the SAM and USFM models which typically cannot be fine-tuned without higher-end GPUs: the encoder of the SAM and USFM models is based on the 86M parameter vit-b and expects images at $1024^2$ (SAM) or $256^2$ (USFM) resolution. This means that our model is a much more modest candidate for downstream fine-tuning, as its backward pass on smaller $128^2$ images will usually fit well into an entry-level GPU with similar or better downstream performance. We publish our code and model weights as a starting point for specialized downstream applications in ultrasound.